\documentclass[preprint2]{aastex}

\newcommand{\myemail}{mahy@astro.ulg.ac.be}
\newcommand{\teff}{\ifmmode T_{\rm eff} \else $T_{\mathrm{eff}}$\fi}
\newcommand{\logg}{\ifmmode \log g \else $\log g$\fi}
\newcommand{\lL}{\ifmmode \log \frac{L}{L_{\odot}} \else $\log \frac{L}{L_{\odot}}$\fi}

\newcommand{\vsini}{$v \sin i$}

\newcommand{\kms}{km s$^{-1}$}
\newcommand{\msun}{\ifmmode M_{\odot} \else M$_{\odot}$\fi}

\shorttitle{A new inverstigation of \object{HD\,48099}}
\shortauthors{Mahy et al.}

\begin{document}

\title{A NEW INVESTIGATION OF THE BINARY \object{HD\,48\,099}}

\author{L. Mahy\altaffilmark{1,9}, G. Rauw\altaffilmark{1,2}, F. Martins\altaffilmark{3}, Y. Naz\'e\altaffilmark{1,2}, E. Gosset\altaffilmark{1,5}, M. De Becker\altaffilmark{1,4}, H. Sana\altaffilmark{6,7}, P. Eenens\altaffilmark{8}}

\altaffiltext{1}{Astrophysical Institute, University of Li\`ege, B\^at. B5C, All\'ee du 6 Ao\^ut 17, B-4000 Li\`ege, Belgium}
\altaffiltext{2}{Research Associate F.R.S.-FNRS}
\altaffiltext{3}{GRAAL, Universit\'e Montpellier II, CNRS, Place Eug\`ene Bataillon, 34095 Montpellier, France}
\altaffiltext{4}{Postdoctoral Researcher F.R.S.-FNRS}
\altaffiltext{5}{Senior Research Associate F.R.S.-FNRS}
\altaffiltext{6}{European Southern Observatory, Alonso de Cordova 1307, Casilla 19001, Santiago 19, Chile}
\altaffiltext{7}{Sterrenkundig Instituut ``Anton Pannekoek'', Universiteit van Amsterdam, Postbus 94249, 1090 GE Amsterdam, The Netherlands}
\altaffiltext{8}{Departamento de Astronomia, Universidad de Guanajuato, Apartado 144, 36000 Guanajuato, GTO, Mexico}
\altaffiltext{9}{\myemail}

\begin{abstract}
With an orbital period of about 3.078 days, the double-lined spectroscopic binary \object{HD\,48\,099} is, still now, the only short-period O$+$O system known in the \object{Mon\,OB2} association. Even though an orbital solution has already been derived for this system, few information are available about the individual stars.\\
We present, in this paper, the results of a long-term spectroscopic campaign. We derive a new orbital solution and apply a disentangling method to recover the mean spectrum of each star. To improve our knowledge concerning both components, we determine their spectral classifications and their projected rotational velocities. We also constrain the main stellar parameters of both stars by using the CMFGEN atmosphere code and provide the wind properties for the primary star through the study of IUE spectra.\\
This investigation reveals that \object{HD\,48\,099} is an O5.5\,V\,((f))\,$+$\,O9\,V binary with $M_1 \sin^3 i = 0.70\,M_{\sun}$ and $M_2 \sin^3 i = 0.39\,M_{\sun}$, implying a rather low orbital inclination. This result, combined with both a large effective temperature and $\log g$, suggests that the primary star (\vsini\,$\simeq$\,91\,km\,s$^{-1}$) is actually a fast rotator with a strongly clumped wind and a nitrogen abundance of about 8 times the solar value.
\end{abstract}


\section{Introduction}

\object{HD\,48\,099} is a double-lined spectroscopic binary situated near the young open cluster NGC\,2244 and embedded in the Monoceros OB2 association in the Rosetta nebula \citep{mah09}. This star has first been quoted as a binary by \citet{sle56} and a period of about 3.1~days was derived by \citet{gar80} and \citet{sti96}. \object{HD\,48\,099} is, still now, the only short-period O$+$O system detected in this association \citep[see e.g.,][]{lin08,mah09}. 

According to the spectroscopic parallax, \object{HD\,48\,099} is located at a distance of 1829$\pm$77\,pc \citep{bb05}, i.e., a distance close to that of NGC\,2244 \citep{hen00}, and seems to be a binary system with a low inclination. Indeed, \citet{gar80} estimated minimum masses of about 0.63 and 0.38\,M$_\sun$ and reported the semi-amplitude of the radial velocity (RV) curves to be 55 and 92\,km\,s$^{-1}$ for the primary and secondary stars, respectively. However, there has been no detailed study of each component. Other stellar parameters were reported only for the global system, e.g., a spectral type of O7\,V \citep{wal72,gcc82}, \teff\,=\,30700$\pm$700\,K, and $\log\,(L/L_{\sun})\,=\,5.2\pm0.1$ \citep{ll06}. 

The present paper is based on the investigation of a series of high-resolution spectra, which help us to improve our knowledge of the main individual stellar parameters of \object{HD\,48\,099}. We applied a disentangling method, using an algorithm inspired by the work of \citet{gl06}, to obtain the mean individual spectrum of each star from the observed data of the entire system. We then derived the spectral classification, the \vsini, and the main stellar parameters for each component.

We thus organize this paper as follows. We first describe our entire dataset and the data reduction in Section \ref{sec:observ}. In Sect.\,\ref{sec:orbital}, the orbital solution and the properties of the disentangled spectra are presented. Section\,\ref{sec:vsini} reports the projected rotational velocities (\vsini)~of both components. We give the spectral type of each star of \object{HD\,48\,099} in Sect.\,\ref{sec:spec}, while Sect.\,\ref{sec:abundance} is devoted to the study of the main stellar parameters by using the model atmosphere code CMFGEN \citep{hm98}. Finally, Sect.\,\ref{sec:concl} discusses the results and provides the conclusions.

\section{Observations and data reduction}
\label{sec:observ}

The majority of our data were taken with the Aur\'elie spectrograph mounted on the 1.52\,m telescope at the Observatoire de Haute-Provence (OHP). The detector used is a 2048\,$\times$\,1024 CCD EEV 42--20 \#3 with a pixel surface of 13.5~$\mu$m$^{2}$. The spectra were obtained with a 600 lines/mm grating, allowing us to reach a resolving power of R\,=\,8\,000 in the blue range (7 spectra centered on 4700\AA~in the [4450--4900]\AA~domain) and R\,=\,10\,000 in the yellow range (15 spectra centered on 5700\AA~in the [5500--5900]\AA~region). Typical exposure times range from 20 to 45 minutes to reach a mean signal-to-noise ratio (SNR) close to 270. The entire dataset was reduced using the \texttt{MIDAS} software developed at the European Southern Observatory (ESO). The wavelength calibration is based on a series of Thorium-Argon comparison spectra, immediately taken before and after the stellar spectrum. The normalization of our data was done by fitting polynomials of degree four to carefully chosen continuum windows.

Another part of our data was obtained with the Fiber-fed Extended Range Optical Spectrograph (FEROS) \'echelle spectrometer mounted on the 2.2\,m ESO/MPG telescope at ESO (La Silla). We collected four spectra in March 2006 (run ESO 076.D-0294(A)) and we retrieved three other spectra from the ESO-archives taken in January and February 2006 (PI: Casassus, run ESO 076.C-0431(A) and PI: Lo Curto, run ESO 076.C-0164(A)). With a resolving power of R\,=\,48\,000 and typical exposure times ranging from 2.5 to 10 minutes, the spectrograph reaches a mean SNR greater than 300 for our target. The data were reduced with a modified FEROS pipeline working under \texttt{MIDAS} environment. In addition to the modifications already described in \citet{san06a}, several new features were implemented as mentioned in \citet{san2009}.

Moreover, the 2.1\,m telescope at the Observatorio Astron\'omico Nacional of San Pedro M\'artir (Mexico) equipped with the Espresso spectrograph allowed us to obtain spectra over five nights in April 2007. This \'echelle spectrograph covers 27 spectral orders over a wavelength domain equal to [3780--6950]\AA~with a resolving power of R\,=\,18\,000. This instrument features a SITE 3 optical CCD chip with 1024\,$\times$\,1024 pixels of 24~$\mu$m$^2$. Consecutive data of a same night, taken with exposure times between 5 to 10 minutes, were added to obtain spectra with a mean SNR close to 250. The data were reduced with the \'echelle package within the MIDAS software. 

Finally, we retrieved two archival Elodie spectra observed by Catala and Bouret in January 2003 and November 2005, respectively. Elodie was an \'echelle spectrograph, which gave 67 spectral orders covering the [3850--6850]\AA~wavelength domain and offering a resolving power of  R\,=\,42\,000. This spectrograph was mounted on the 1.93\,m telescope at OHP and exposure times of 12 and 33 minutes provided a SNR of about 130 and 100, respectively.\\ The journal of the observations presenting the 36 spectra is listed in Table~\ref{table1:overview}.

\section{Orbital solution}
\label{sec:orbital}

The analysis of \citet{gar80}, based on an orbital period of about 3.1~days, allowed to compute, for the first time, the minimum masses and semi-amplitude of the RV curves of \object{HD\,48\,099}. However, the first complete orbital solution, available in the literature, was derived by \citet{sti96}. He used an orbital period of 3.07809~days and assumed a zero eccentricity to derive, from 23 high-resolution IUE spectra (SWP data), the main orbital parameters of \object{HD\,48\,099}.

To determine the orbital motion of the primary and secondary components, we proceed in two steps. To obtain a first approximation of the RVs of both stars, we measured these values by fitting Gaussians to the spectral line profiles of the spectra where the two components were deblended. To this aim, we focus on the \ion {Si}{4}\,$\lambda$\,4089, \ion{He}{1}\,$\lambda$\,4471, \ion{Mg}{2}\,$\lambda$\,4481, and \ion{He}{1}\,$\lambda$\,5876 lines. We note that the rest wavelengths of these lines were taken from the papers of \citet{con77} and \citet{und94}. 

Then, a disentangling method, based on the approach of \citet{gl06}, is applied to separate, in an iterative procedure, the two components of the binary system. This consists of alternately using the spectrum of one component (shifted by its radial velocity) to remove it from the observed spectra in order to calculate a mean spectrum of the other component. Our algorithm also uses the cross-correlation technique to compute the RVs from the disentangled spectra, even at phases for which lines are heavily blended. This method thus allows us to determine RVs for the blended spectra, while the fit with Gaussian functions only allows us to measure the RVs for the deblended spectra. For the purpose of building the RV cross-correlation masks, we used a common basis of spectral lines including the \ion{He}{1}\,$\lambda$\,4471, \ion{Mg}{2}\,$\lambda$\,4481, \ion{He}{1}\,$\lambda$\,4713, \ion{O}{3}\,$\lambda$\,5592, and \ion{He}{1}\,$\lambda$\,5876 lines. We added the \ion{He}{2}\,$\lambda$\,4542, \ion{He}{2}\,$\lambda$\,4686, \ion{C}{4}\,$\lambda$\,5801, and \ion{C}{4}\,$\lambda$\,5812 lines for the primary, and \ion{N}{3}\,$\lambda$\,4511 and \ion{N}{3}\,$\lambda$\,4515 for the secondary. The RVs are listed in Table~\ref{table1:overview}. To avoid degrading the best quality data, we have thus interpolated the spectra with a poor resolution to obtain a step similar to the FEROS data. Since the bulk of our data covers the blue and yellow range, we favor the disentangling on two wavelength domains: [4450--4900]\AA~and [5500--5920]\AA. We note that the first region is particularly useful to establish an accurate determination of the spectral classification of each component of the binary system. We also emphasize that the spectra have been disentangled in the [4000--4220]\AA~wavelength domain but because of the poor time sampling of the data in this domain, the result should only be considered as preliminary.

\begin{figure}[htbp]
  \centering
  \includegraphics[width=7cm, bb=48 160 570 705, clip]{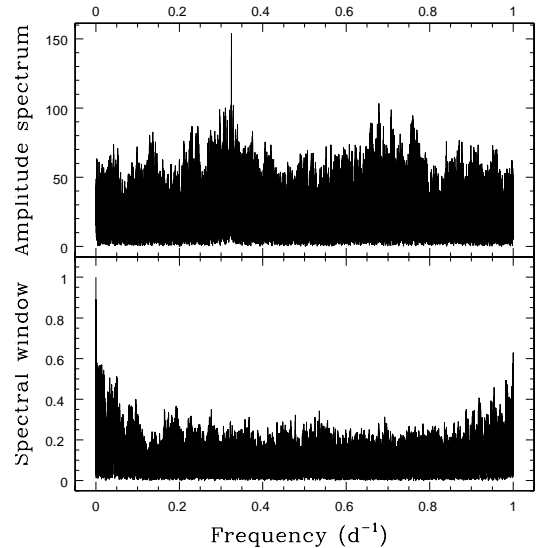}
  \caption{{\it Top:} Semi-amplitude Fourier spectrum computed from our new RVs along with the RVs measured from the IUE-archive spectra by \citet{sti96}.
    {\it Bottom:} Spectral window corresponding to the sampling of the entire (our + IUE) RV dataset.}
  \label{powerspectrum}
\end{figure}

To compute the improved orbital solution of \object{HD\,48\,099}, we first refined the orbital period by combining the RVs of \citet{sti96}, after having shifted these values to obtain the same systemic velocity as for our data, with the RVs calculated in the present paper. Using these two datasets, we cover a total time interval of $\Delta T = 10684.4$~days, implying a natural peak width in the periodogram equal to $1/\Delta T = 9.3593~10^{-5}~d^{-1}$. We then apply a Fourier method that is especially designed for astrophysical time series with a highly non-uniform time sampling \citep[see][]{hec85,gos01} and which was successfully applied to a number of systems studied by our team. The semi-amplitude spectrum (Fig. \ref{powerspectrum}), evaluated on the RV$_P$$-$RV$_S$ time series by the Fourier analysis, presents a strong peak at the frequency of 0.324880\,d$^{-1}$ which corresponds to an orbital period of 3.07806\,$\pm$\,0.00009 days. The error quoted on the period is assumed to be equal to one tenth of the width of the peaks of the periodogram. We have rejected the hypothesis that the highest peak is due to white noise, by taking a threshold significance level of 0.01.

\begin{figure}[htbp]
  \centering
  \includegraphics[width=7.5cm, bb=20 170 550 690, clip]{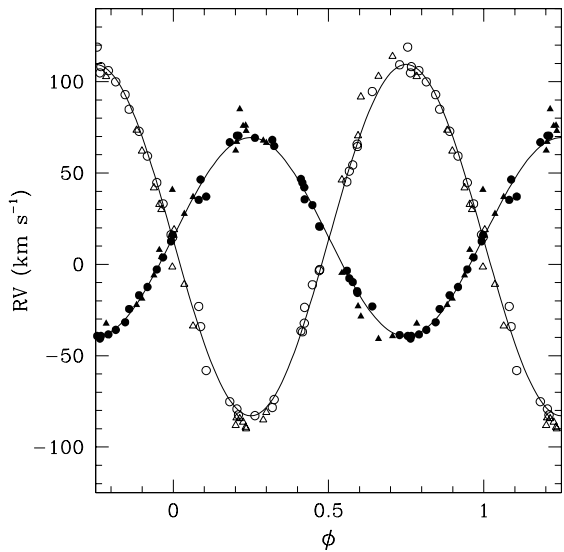}
  \caption{Radial velocity curve of HD\,48\,099 computed only with our RVs measured by fitting Gaussians for the deblended spectra and by cross-correlation for the blended ones. The primary data correspond to the filled symbols and the open ones indicate the radial velocities of the secondary component. Our spectra are represented with circles. We added (by triangles), for comparison, the RVs measured by \citet{sti96} but shifted to obtain the same systemic velocity as for our data.}
  \label{curve}
\end{figure}

Adopting this period, we used the LOSP (Li\`ege Orbital Solution Package) program on our data to compute the orbital solution of \object{HD\,48\,099}. This software is based on the generalization of the SB1 method of \citet{wol67} to the SB2 case along the lines described in \citet{rau00} and \citet{san06a}; more details as well as the present status of the software can be found in Sana\,\&\,Gosset (A\&A, submitted). We assigned a weight of 1.0 to all spectra of our dataset independently of the primary or secondary star. Fig. \ref{curve} represents the fitted RV curves as a function of the orbital phase and Table \ref{table2:param} yields the orbital parameters computed for this star with their corresponding uncertainties. The fit residuals are smaller with a zero eccentricity indicating that the circularization of the orbit is achieved. Although the results obtained are close to those of \citet{sti96}, we however underline significant differences notably on the minimal masses of both components. We also emphasize that our orbital solution is subject to smaller uncertainties.

\begin{figure}[htbp]
  \centering
  \includegraphics[width=7cm, bb=22 154 582 710, clip]{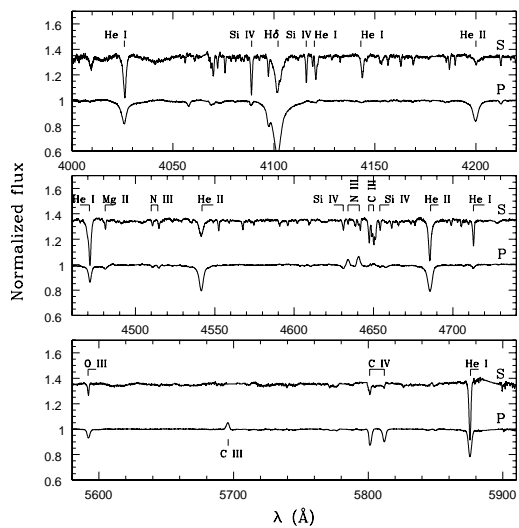}
  \caption{Normalized disentangled spectra of HD\,48\,099 between [4000--4220]\AA, [4450--4730]\AA, and [5580--5900]\AA. In each panel, the lower spectrum is that of the primary. The secondary spectrum is vertically shifted by 0.4 units for clarity.}
  \label{disent}
\end{figure}

The RVs corresponding to the observed RV curves are then fixed in the input file of the disentangling program to derive the individual mean spectra presented in Fig. \ref{disent}. Because of some uncertainties on the continuum levels in the observed input spectra, this disentangling technique generates some low level, low frequency oscillations ($\sim 5\%$ of the continuum) in the normalization of the disentangled spectra (see e.g., $\lambda$~5890 in Fig.~\ref{disent}) but such features have no impact on our scientific results.

Finally, we also used the disentangling method to separate the IUE spectra. We focus on the [1210--1780]\AA~wavelength domain. This region features many spectral lines. On the basis of the orbital solution, we determined the phase corresponding to each observed IUE spectrum. The theoretical RV curves then gave us the associated RVs for the different phases. As we have done for the optical spectra, we fixed these values in the input file of the program to obtain the mean IUE spectrum of both components. However, the disentangled UV spectrum of the secondary is too noisy to allow us to study the wind properties of this star (see Sect.~\ref{sec:abundance}).

\section{Projected rotational velocities}
\label{sec:vsini}

A first estimate of the \vsini~was derived by \citet{how97} to be about 81 and 41 km~s$^{-1}$ for the primary and secondary stars, respectively.

\begin{figure}[htbp]
  \centering
  \includegraphics[width=9cm, bb=54 203 575 692, clip]{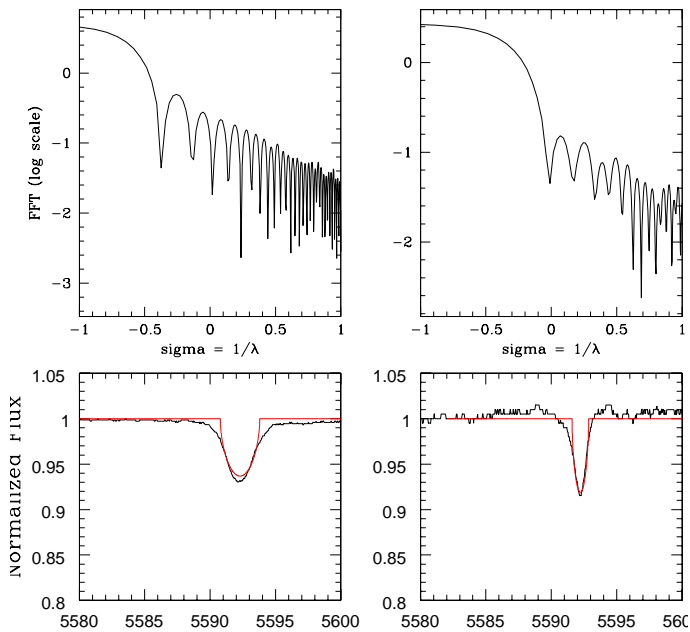}
  \caption{{\it Top:} Plots of the Fourier transform computed for the \ion{O}{3}~$\lambda$~5592 line of the primary star ({\it left}) and the secondary component ({\it right}). {\it Bottom:} Synthetic rotational profiles measured for the same lines without taking into account the macroturbulence.}
  \label{rotation}
\end{figure}

The \vsini~is computed for each component by using the Fourier transform method \citep{sim07} based on a linear limb-darkening law \citep[for detailed explanations and coefficients, see][]{gray05}. For the primary, the mean \vsini, computed from the profiles of the \ion{He}{1}\,$\lambda$\,4471, \ion{He}{2}\,$\lambda$\,4542, \ion{O}{3}\,$\lambda$\,5592 (Fig.~\ref{rotation} {\it left}), \ion{C}{4}\,$\lambda$\,5812 and \ion{He}{1}\,$\lambda$\,5876 lines, is equal to 91\,$\pm$\,12\,km$\,$s$^{-1}$. For the secondary, we take into account the \ion{He}{1}\,$\lambda$\,4471, \ion{He}{2}\,$\lambda$\,4542, \ion{He}{1}\,$\lambda$\,4713, \ion{O}{3}\,$\lambda$\,5592 (Fig.~\ref{rotation} {\it right}) and \ion{He}{1}\,$\lambda$\,5876 lines since \ion{C}{4}\,$\lambda$\,5812 appears very weakly in the mean spectrum. We derive a mean value for \vsini~of 51\,$\pm$\,17\,km$\,$s$^{-1}$. 

\section{Spectral classifications}
\label{sec:spec}

First, we have measured the equivalent widths ($EW$s) by fitting two Gaussians to the line profiles of observed spectra corresponding to phases close to the quadratures. The spectral type determination is based on the quantitative classification criteria for O-type stars of \citet{con71}, \citet{con73} and \citet{mat88,mat89}. In consequence, we adopt the usual notations: $\log(W') = \log$($EW$\,\ion{He}{1}\,$\lambda$\,4471/$EW$\,\ion{He}{2}\,$\lambda$\,4542), $\log(W'') = \log$($EW$\,\ion{Si}{4}\,$\lambda$\,4089/$EW$\,\ion{He}{1}\,$\lambda$\,4143), $\log(W''') = \log$($EW$\,\ion{He}{1}\,$\lambda$\,4388) + $\log$($EW$\,\ion{He}{2}\,$\lambda$\,4686). We find $\log(W') = -0.31 \pm 0.02$ and $\log(W') = 0.40 \pm 0.01$ for the primary and the secondary, respectively. We then compared those values to the ones measured on the disentangled spectra, finding that they are in good agreement with each other (see Table\,\ref{ewtable}). On the one hand, the spectral type of the primary star is thus intermediate between two spectral classifications: O5.5 and O6. However, the O5.5 classification is favored after a comparison with the spectra in the atlas of \citet{wal90}. Moreover, the mean spectrum of the primary component displays a strong absorption for the \ion{He}{2}\,$\lambda$\,4686 line whilst the \ion{N}{3}\,$\lambda \lambda$\,4634$-$4641 lines are in emission, supporting an ((f)) tag. On the other hand, the spectrum of the secondary clearly reveals a late O-type star. The value computed from the ratio between \ion{He}{1}\,$\lambda$\,4471 and \ion{He}{2}\,$\lambda$\,4542 corresponds to an O9 star. With $\log(W'')\,\simeq -0.04$, computed from the \'echelle spectra, Conti's criteria clearly indicate a main-sequence luminosity (V). Furthermore, Mathys O8--O9.7 luminosity criterion confirms that the secondary component lies on the MS band ($\log(W''')\,\simeq\,5.47$). In summary, we assign an O5.5\,V\,((f)) spectral type for the primary and an O9\,V one for the secondary.

We then estimated the brightness ratio $l_1$/$l_2$$\,=\,3.96\,\pm\,0.21$. Indeed, from the \ion{Si}{4}\,$\lambda$\,4089, \ion{He}{1}\,$\lambda$\,4471, \ion{O}{3}\,$\lambda$\,5592, and \ion{He}{1}\,$\lambda$\,5876 lines, we computed the $EW$ ratio between the primary and secondary, respectively, and the average values calculated from \citet{con71} and \citet{con73} for stars with the same spectral type (for more details, see e.g., \citealt{lin08}). The primary star is thus about four times brighter than the secondary component in the optical domain. This ratio greatly differs from the one computed in the UV domain where \citet{sti96} found a value of 1.8. The discrepancy is surprizing since the spectral type that we have found would suggest that the brightness ratio in the UV domain should be even larger than in the optical. 

The Reed Catalogue \citep{ree05} provides the UBV photometry parameters required to compute the bolometric luminosity of each star. We assumed a combined spectral type O7\,V for the binary system as already mentioned by \citet{wal72} or \citet{gcc82}. In consequence, we used $V = 6.37\,\pm\,0.01$ and $B-V = -0.047\,\pm\,0.010$. Under this assumption and using $(B-V)_0 = -0.27$ \citep{mar06}, we found $E(B-V) = 0.223\,\pm\,0.010$.

Assuming that the distance of 1829$\,\pm\,$77 pc \citep{bb05}, computed for \object{HD\,48\,099} from the spectroscopic parallax, is correct, the $DM$ (distance modulus) of this star is estimated at $11.31\,\pm\,0.09$. The visual absolute magnitude of the entire system would be $M_V = -5.63\,\pm\,0.11$. We thus derive, from the luminosity ratio, $M_{V_P} = -5.39\,\pm\,0.12$ and $M_{V_S} = -3.89\,\pm\,0.06$ for the primary and secondary, respectively. Using the bolometric corrections of \citet{mar06}, we inferred a bolometric magnitude of $-9.09\,\pm\,0.12$ and $-7.02\,\pm\,0.06$ for the primary and secondary stars, respectively, equivalent to absolute luminosities of $\log(L_1/L_{\sun}) = 5.54\,\pm\,0.12$ and $\log(L_2/L_{\sun}) = 4.71\,\pm\,0.06$. The luminosities of both stars are in good agreement with our spectral modelling (see Sect.\,\ref{sec:abundance}), clearly indicating main-sequence stars.  

\section{Modelling spectra for an abundance study}
\label{sec:abundance}

\begin{figure}[htbp]
\epsscale{1}
\begin{center}
\includegraphics[width=8cm,height=8cm]{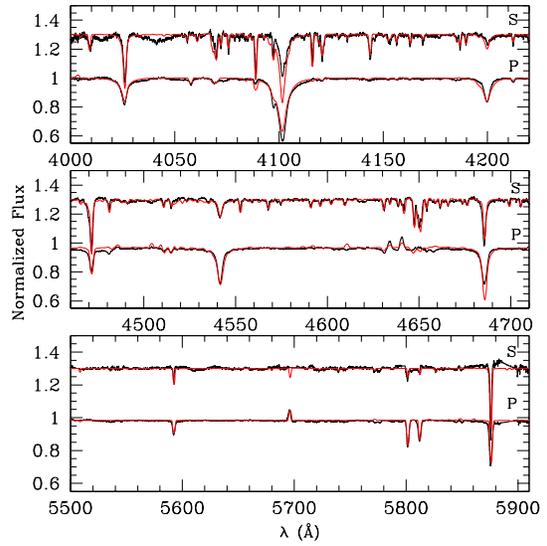}
\end{center}
\caption{Comparison between the disentangled (black line) and the theoretical (red line) optical spectra of HD\,48\,099. The latter were computed with the CMFGEN atmosphere code. The secondary spectrum was vertically shifted by 0.3 units for clarity. }
\label{fig_opt}
\end{figure}

\begin{figure}[htbp]
\epsscale{1}
\begin{center}
\includegraphics[width=8cm,height=8cm]{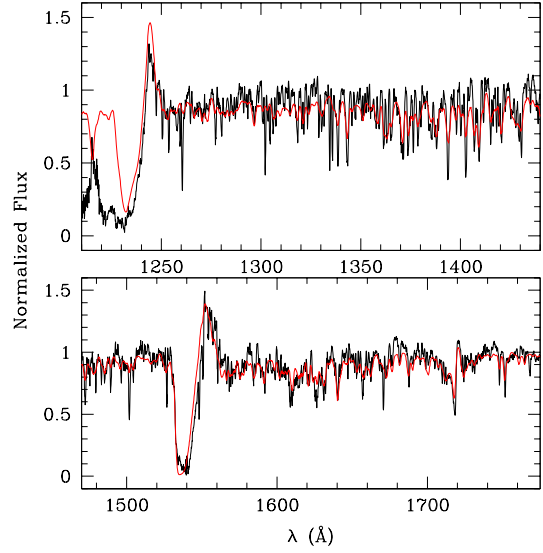}
\end{center}
\caption{Same as Fig.\,\ref{fig_opt} but only for the primary star in the UV domain. The disentangled spectrum was built from the IUE-archive spectra of HD\,48\,099.}
\label{fig_uv}
\end{figure}

We have conducted a spectroscopic analysis of the components of \object{HD\,48\,099} using the code CMFGEN \citep[see][for details]{hm98}. Non-LTE atmosphere models including winds and line--blanketing have been computed and synthetic spectra have been fitted to the disentangled spectra of \object{HD\,48\,099}. We have thus constrained the main stellar parameters of both the primary and the secondary: \teff, \logg, luminosity, and surface N content. We note that the oscillator strengths for the \ion{N}{3} were taken from the atomic data taken from the OPACITY project\footnote{The oscillator strengths were given in atomic files available at the URL: \texttt{http://cdsweb.u-strasbg.fr/topbase/op.html}. The oscillator strengths~($f$) for the relevant \ion{N}{3} lines are: 0.17338 ($\lambda$~4510.88), 0.21845 ($\lambda$~4510.96), 0.27716 ($\lambda$~4514.85) and 0.17310 ($\lambda$~4518.14).}. The abundances of other chemical elements have been fixed to solar since we lack reliable indicators to allow us to change these abundances. The solar abundances of \citet{gas07} have been used unless stated otherwise. For the primary, we have also fitted the disentangled mean IUE spectrum to determine the terminal velocity, mass loss rate, and clumping factor of the stellar wind. Figures\,\ref{fig_opt} and \ref{fig_uv} show the comparison between the disentangled and the theoretical spectra in the optical and UV domains, respectively. The agreement between the model and the observations is rather good for both components of the system. As mentioned in Sect.~\ref{sec:orbital}, the analysis of the individual spectrum of the secondary star in the UV domain is not possible because of its large noise level. 

In practice, we have used the \ion{He}{1}\,$\lambda$\,4471 and \ion{He}{2}\,$\lambda$\,4542 lines to estimate \teff. The gravity was determined from the Balmer lines H$\beta$, H$\gamma$ and H$\delta$. The luminosity was constrained by the V-band magnitude: for the adopted distance and extinction (see Sect.\,5), we computed the V-band magnitude of our synthetic spectra and adjusted the luminosity to correctly reproduce the observed V magnitude. The results are gathered in Table\,\ref{tab_param}. The primary has a high effective temperature, larger than expected for its spectral type \citep{mar05}, but it also has a quite large \logg. We will see below that this could mean either that the primary is a very young O star, or that it is a fast rotator. The secondary has \teff\ and \logg\ more typical for its spectral type. To obtain a good agreement with the primary spectrum, it is important that the model is enriched in nitrogen. The nitrogen abundance is found to be 5 to 10 times the solar value, with the best model showing a nitrogen content equal to 8 times this value. However, we note that the nitrogen content is sensitive to the variation of the effective temperature and the microturbulence (equal to 10 km~s$^{-1}$ in our model). For example, a variation of 1000~K or 10 km~s$^{-1}$ implies a change of about 20\% of the nitrogen abundance.

The wind properties of the primary were constrained from the disentangled IUE spectrum. The \ion{N}{5}\,$\lambda$\,1240, \ion{C}{4}\,$\lambda$\,1550 and \ion{N}{4}\,$\lambda$\,1720 lines were the main indicators. A terminal velocity of 2800 \kms\ convincingly reproduced the blueward extension of the main P--Cygni profiles. We found that a strongly clumped wind (with a filling factor of 0.01) was necessary to correctly reproduce the \ion{N}{4} $\lambda$ 1720 line \citep[see][]{bouret05}. The corresponding mass loss rate
required to fit all P--Cygni lines is 2.5~$10^{-8}$~M$_{\odot}$~yr$^{-1}$. 

\begin{figure}[htbp]
\epsscale{1}
\begin{center}
\includegraphics[width=8cm,height=8cm]{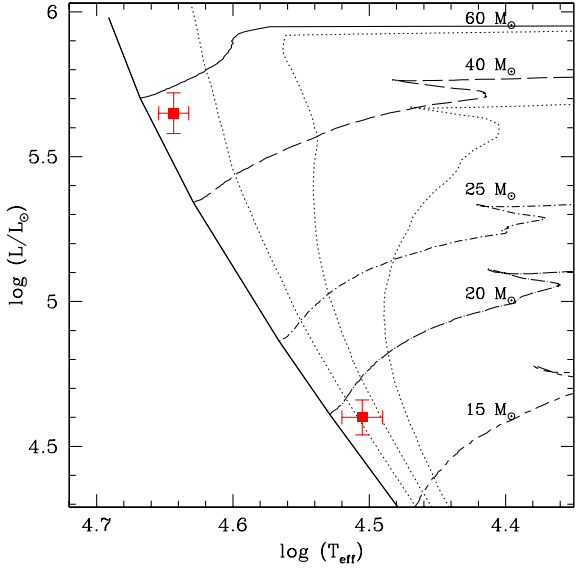}
\end{center}
\caption{HR diagram with the position of the two components of HD\,48\,099 indicated. Evolutionary tracks are from \citet{mm03} for an initial rotational velocity of about 300~\kms. The isochrones shown by dotted lines are for 2, 4 and 6 Myr. }
\label{hrd}
\end{figure}

Figure\,\ref{hrd} shows the position of the two components in a HR diagram built with the Geneva single star evolutionary tracks of \citet{mm03}. Provided that the components of HD\,48\,099 have not gone through an evolutionary phase of mass exchange or mass loss through Roche lobe overflow, the comparison with these evolutionary tracks can provide some indications on the age and absolute parameters of the stars. The secondary appears to be 1--5 Myr old. Its evolutionary mass is 19$\pm$2\,\msun. The primary lies rather close to the main-sequence, indicating an age of 0--1 Myr. A mass of 55$\pm$5 \msun\ is derived from interpolation between tracks. Its effective temperature (44\,kK) is much hotter than what is expected for a classical main-sequence O5.5 star, but its gravity is also quite high \citep[\logg\ $>$~4.5 compared to $\sim$4.0 for class V O stars, see][]{mar05}. The star might thus be younger than normal main-sequence stars. Consequently, it might be hotter and more compact, explaining the larger gravity. However, since the secondary appears rather similar to normal late O dwarfs, we might speculate instead that the evolution of the primary was affected by binarity. The rather large surface nitrogen content (about 8 times solar) is also difficult to explain if the star is very young, while it might be more easily accounted for by binarity. In a binary system, the effect of rotation on the evolution is actually enhanced through the exchange of orbital and rotational angular momentum \citep{dem09}. The latter effects have a strong impact on the determination of the stellar parameters and a comparison with single star tracks might not be valid anymore. This idea remains to be tested by theoretical (evolutionary) models of binary evolution. Another possibility to explain the properties of the primary would be fast rotation. Since the system is seen at low inclination, and assuming that the primary and secondary have their spin axes aligned with the system's angular momentum vector, we cannot exclude a large rotational velocity for the primary (its \vsini\ is $\sim$1.8 times that of the secondary). Fast rotating single stars tend to evolve more ``vertically'' than normal stars in the early phases of evolution \citep[see a discussion of this effect in][]{mar09}. This could explain the larger \teff\ and enhanced N/H.

\section{Discussion and conclusions}
\label{sec:concl}

Our detailed investigation of the short-period binary HD\,48\,099 revealed a system composed of an O5.5\,V\,((f)) primary and an O9\,V secondary. We have studied the main stellar parameters of both stars by using, on the one hand, the CMFGEN code to fit the optical spectra, and, on the other hand, the visual absolute luminosity of both components. These two approaches provided quite similar luminosities. We thus adopt for \object{HD\,48\,099} the values given in Sect.\,\ref{sec:abundance} i.e., $\log(L_1/L_{\sun}) = 5.65\,\pm\,0.07$ and $\log(L_2/L_{\sun}) = 4.60\,\pm\,0.06$. However, a significant difference remains in the mass ratio of the system. Indeed, the RVs  yield a value of about 1.77, while the evolutionary tracks indicate a value close to 2.89. The discrepancy could be seen as an indication that the primary is an evolved star that has lost part of its mass through a Roche lobe overflow or it could again be due to fast rotation (see the discussion in Sect.\,\ref{sec:abundance}).

The orbital solution, computed in the present paper, is particularly marked by the very small minimum masses and the low amplitudes of the RV curves. As \citet{sti96} already suggested, the orbit of the binary system should thus be seen under a low inclination angle. From the theoretical estimates proposed by \citet{mar05}, the gravitational masses of an O5.5\,V and an O9\,V star correspond to about 34 M$_{\sun}$ and 18 M$_{\sun}$, respectively. Under this assumption, we can estimate the inclination of the \object{HD\,48\,099} binary system to be about 16$^{\circ}$. We note that a modification of the masses corresponding to $\pm$1~spectral type alters the inclination of the system by about 1$^{\circ}$ whilst the use of evolutionary masses given in Table~\ref{tab_param} leads to a change of the inclination by about 2$^{\circ}$. However, we must be careful with the comparison between the stellar parameters given in Table~\ref{tab_param} for both stars of HD\,48099 and the values quoted in \citet{mar05} for a same spectral type since the latter were given for \logg~$ = 3.92$. The \teff, $L$ and the masses could be directly affected by the difference of \logg~found for the two stars in the binary system and could explain why the \teff~found by CMFGEN is too high compared with the values given in \citet{mar05}.

By adopting an inclination of 16$^{\circ}$, the rotational velocity of both components is close to 330 and 185\,km\,s$^{-1}$ for the primary and secondary, respectively. \object{HD\,48\,099} could thus contain a rapid rotator. Such an assumption could explain why the primary star appears more massive in the evolutionary track than in reality \citep{mm03,mar09}. In addition, by assuming that the inclination of the rotation axis of both components is the same as the orbital inclination ($\sim 16^{\circ}$), the stars would essentially be seen pole-on. Since the fast rotation of the primary would make this star hotter near the poles than at the equator, the spectral classification for this star could be biased towards earlier types.

Taking into account the radii of both stars (see Table\,\ref{tab_param}), the rotational velocities suggest that the two components are in synchronous rotation, i.e. the angular velocity is similar for both stars, but exclude a system in synchronous co-rotation, i.e. the orbital period of the system is not the same as the rotation period of both stars. Indeed, to obtain an angular velocity of the system equal to the individual angular rotation velocities of the stars, we must assume that the inclination of the orbit is equal to 28.5$^{\circ}$, implying too small masses for such a system. On the basis of an inclination of $16^{\circ}$ for both stars, the Roche lobe radii are equal to 14.3 and 11.0~$R_{\sun}$ for the primary and secondary, respectively. The ratio between the stellar and Roche lobe volumes indicate that the primary star, with a radius of 11.6~$R_{\sun}$, fills its Roche lobe at 53\%, whilst the secondary only occupies 21\%. This result thus suggests that, for the young system \object{HD\,48\,099}, the fast rotation could be the main factor of evolution. However, we note that a relatively small error on the inclination could have a large impact on the evolutionary status (Roche lobe overflow) of HD\,48\,099 since, by taking an inclination of about 20$^{\circ}$, the primary star would fill its Roche lobe. 

Finally, we put in perspective the two O$+$O binary systems known in Mon\,OB2 association (HD\,48\,099 and Plaskett's star). The comparison shows that the most massive star of both system is affected by nitrogen enrichment. We derived a N overabundance for the fast rotator in HD\,48\,099, while \citet{lin08} reported, for Plaskett's star, that the componenet with the slowest rotation velocity has a N content close to 16 times the solar abundance. However, we emphasize that Plaskett's star is in a post case A Roche lobe overflow stage.

\acknowledgments
This research is supported by the FNRS (Belgium), by a PRODEX XMM/Integral contract (Belspo) and by the Communaut\'e fran\c caise de Belgique - Action de recherche concert\'ee (ARC) - Acad\'emie Wallonie--Europe. We acknowledge the Minist\`ere de l'Enseignement Sup\'erieur et de la Recherche de la Communaut\'e Fran\c caise for supporting our travels to OHP. We also thank the staff of Observatoire de Haute-Provence and of La Silla ESO Observatory for their technical support. F.M. thanks John Hillier for making his code available and for assistance. P.E. acknowledges support through CONACyT grant 67041.

\bibliographystyle{apj}
\bibliography{laurent}

\newpage

\begin{table}[htbp]
\begin{center}
\caption{Journal of the observations of HD\,48\,099. The first column gives the name of the instrument with which the spectrum has been acquired. The second one lists the heliocentric Julian date (HJD), and the third provides the phase computed from our orbital solution (see Table\,\ref{table2:param}). The last two columns give the measured radial velocities (in km\,s$^{-1}$) of the primary and secondary stars, respectively, at the indicated phase.\label{table1:overview}}
\begin{tabular}{l c c r r}
\tableline\tableline
Instrument & HJD & $\Phi$ & RV$_P$ & RV$_S$\\
& -- 2\,450\,000 & & &\\
\tableline
Elodie & 2655.3793 & 0.863 & -24.4 &  84.9\\
Elodie & 3682.5907 & 0.580 &  -9.7 &  54.4\\
FEROS  & 3738.7235 & 0.816 & -35.9 &  99.9\\
FEROS  & 3774.5994 & 0.471 &  20.8 &  -3.5\\
FEROS  & 3774.6070 & 0.474 &  20.6 &  -2.8\\
Aur\'elie& 3775.5035 & 0.765 & -40.7 & 104.8\\
Aur\'elie& 3775.5146 & 0.769 & -39.2 & 108.2\\
Aur\'elie& 3776.5014 & 0.089 &  46.5 & -34.1\\
Aur\'elie& 3777.4952 & 0.412 &  46.7 & -36.4\\
Aur\'elie& 3778.4762 & 0.731 & -38.7 & 109.3\\
Aur\'elie& 3779.2880 & 0.995 &  12.6 &  16.3\\
Aur\'elie& 3779.3019 & 0.999 &  16.1 &  14.9\\
Aur\'elie& 3780.3088 & 0.326 &  64.8 & -74.0\\
FEROS  & 3796.5248 & 0.595 & -15.6 &  64.5\\
FEROS  & 3797.5211 & 0.918 & -12.4 &  59.3\\
FEROS  & 3799.5214 & 0.568 &  -7.6 &  51.0\\
FEROS  & 3800.5172 & 0.891 & -16.9 &  72.8\\
Espresso& 4193.7415& 0.641 & -23.0 &  94.6\\
Espresso& 4194.7463& 0.967 &   3.8 &  33.1\\
Espresso& 4197.7630& 0.947 &  -2.8 &  44.8\\
Espresso& 4198.7354& 0.263 &  69.3 & -82.8\\
Espresso& 4199.7469& 0.592 & -14.7 &  65.8\\
Aur\'elie& 4200.3600 & 0.791 & -38.4 & 106.0\\
Aur\'elie& 4205.3612 & 0.416 &  44.6 & -37.0\\
Aur\'elie& 4205.3793 & 0.422 &  42.2 & -32.2\\
Aur\'elie& 4401.6362 & 0.181 &  66.9 & -75.2\\
Aur\'elie& 4410.6362 & 0.105 &  37.1 & -58.1\\
Aur\'elie& 4411.6912 & 0.448 &  32.4 & -11.2\\
Aur\'elie& 4412.6364 & 0.755 & -39.2 & 118.9\\
Aur\'elie& 4413.6435 & 0.081 &  35.3 & -23.1\\
Aur\'elie& 4414.6915 & 0.422 &  35.6 & -23.6\\
Aur\'elie& 4423.6053 & 0.318 &  68.1 & -78.4\\
Aur\'elie& 4472.5179 & 0.209 &  70.4 & -82.3\\
Aur\'elie& 4473.5943 & 0.558 &  -3.5 &  45.2\\
Aur\'elie& 4474.4721 & 0.844 & -31.7 &  92.9\\
Aur\'elie& 4475.5804 & 0.204 &  70.5 & -79.2\\
\tableline
\end{tabular}
\end{center}
\end{table}

\begin{table}[htbp]
\begin{center}
\caption{Orbital solution for HD\,48\,099 computed by assuming that the uncertainties on the secondary RVs are twice as large as for the primary. $T_0$ refers to the time of the primary conjunction. $\gamma$, $K$, $a\,\sin\,i$ correspond to the apparent systemic velocity, the semi-amplitude of the RV curve, and the projected semi-major axis, respectively. \label{table2:param}}
\begin{tabular}{l c}
\tableline\tableline
Parameters & Primary \qquad Secondary\\
\tableline
$P$(days) & 3.07806 $\pm$ 0.00009\\
$e$& 0.0 (fixed)\\
$T_0$ (HJD)& 2$\;$452$\;$649.661 $\pm$ 0.004\\
$\gamma$ (km\,s$^{-1}$)& 15.0 $\pm$ 0.6 \qquad 13.3 $\pm$ 0.8\\
$K$ (km\,s$^{-1}$)& 54.4 $\pm$ 0.7 \qquad 96.2 $\pm$ 1.2\\
$a\,\sin\,i$ ($R_{\sun}$) & 3.31 $\pm$ 0.04 \qquad \,5.85 $\pm$ 0.08\\
$M\,\sin^3\,i$ ($M_{\sun}$) & 0.70 $\pm$ 0.02 \qquad \,0.39 $\pm$ 0.01\\
$Q$ ($M_1/M_2$)& 1.77 $\pm$ 0.03\\
rms (km\,s$^{-1}$) & 2.71\\
\tableline
\end{tabular}
\end{center}
\end{table}

\begin{table*}[htbp]
\begin{center}
\caption{Equivalent widths measured directly from some observed spectra. The first column represents the HJD$-$2\,450\,000, the second one gives the orbital phase derived from the orbital solution. The next columns report the $EW$s measured from the data (expressed in m\AA), along with the $\log(W')$ and $\log(W'')$ quantities as defined in the text. The measured $EW$s were corrected from the brightness ratio ($l_1/l_2¸=¸3.96$).\label{ewtable}}
\begin{tabular}{l|c|c c|c c|c c|c c|c c|c}
\tableline\tableline
HJD & $\Phi$ & \multicolumn{2}{c|}{$EW$$_{4471}$} & \multicolumn{2}{c|}{$EW$$_{4542}$} & \multicolumn{2}{c|}{$\log(W')$} & \multicolumn{2}{c|}{$EW$$_{4089}$} & \multicolumn{2}{c|}{$EW$$_{4143}$} & $\log(W'')$ \\
$-$~2\,450\,000   &        & P & S & P & S & P & S& P & S & P & S & \\
\tableline
2655.3793 & 0.863 & 332 & 600 & 714 & 238& $-$0.33 & 0.40 &$-$&$-$ & $-$&$-$& $-$\\
3738.7235 & 0.816 & 372 & 684 & 740 & 268& $-$0.30 & 0.41 &41 & 322& 39 & 377& $-$0.07\\
3796.5248 & 0.595 & 379 & 551 & 754 & 208& $-$0.30 & 0.42 &44 & 352& 36 & 387& $-$0.04\\
3797.5211 & 0.918 & 371 & 600 & 779 & 243& $-$0.32 & 0.39 &39 & 347& 44 & 427& $-$0.09\\
4401.6362 & 0.181 & 357 & 615 & 773 & 243& $-$0.34 & 0.40 &$-$&$-$ & $-$&$-$& $-$\\
4412.6364 & 0.755 & 351 & 784 & 705 & 312& $-$0.30 & 0.40 &$-$&$-$ & $-$&$-$& $-$\\ 
4193.7415 & 0.641 & 391 & 625 & 757 & 253 & $-$0.29 & 0.39 & $-$ & $-$ & $-$ & $-$ & $-$\\
\tableline
disentangled &$-$ & 338 & 761& 711 & 322 & $-$0.32 & 0.37 & 65 & 267 & 70 & 254 & 0.02\\
\tableline
\end{tabular}
\end{center}
\end{table*}

\begin{table}[htbp]
\begin{center}
\caption{Summary of the main chemical elements, super-levels and levels used in the models. We also take into account \ion{Ne}{2}$-${\small{IV}}, \ion{S}{4}$-${\small{V}}, \ion{Ar}{3}$-${\small{V}}, \ion{Ca}{3}$-${\small{IV}} and \ion{Ni}{3}$-${\small{VI}} species. \label{species}}
\begin{tabular}{l r r|l r r|l r r}
\tableline\tableline
Species & super-& levels & Species & super-& levels & Species & super-& levels \\
&levels & & & levels & & & levels & \\
\tableline
\ion{H}{1}& 30 & 30 & \ion{N}{4} & 44& 70& \ion{Si}{4} & 66 & 66\\
\ion{He}{1} & 69 & 69& \ion{N}{5} & 41 & 49& \ion{Fe}{3} & 65 & 607\\
\ion{He}{2} & 30 & 30&\ion{O}{3}& 30 &75& \ion{Fe}{4}& 100 & 1000\\
\ion{C}{3}& 99 & 243&\ion{O}{4} & 30 & 64&\ion{Fe}{5}& 139 & 1000\\
\ion{C}{4} & 64 & 64& \ion{O}{5} & 32 & 56& \ion{Fe}{6}& 59 & 1000\\
\ion{N}{3}&57 & 287& \ion{Si}{3} & 50 & 50&\ion{Fe}{7}& 40 & 245\\
\tableline
\end{tabular}
\end{center}
\end{table}

\begin{table}
\begin{center}
\caption{Derived stellar parameters for the two components of HD\,48\,099. The masses are from interpolation in the single star evolutionary tracks. \label{tab_param}}
\begin{tabular}{lrrrrrrrrrr}
\hline\hline
Parameter  & Primary & Secondary \\
\hline   
\teff\ (kK)   & 44$\pm$1      & 32$\pm$1  \\
\logg\       & 4.5$\pm$0.1    & 3.5$\pm$0.1  \\
\lL\         & 5.65$\pm$0.07  & 4.60$\pm$0.06 \\
$M_{\textrm{theo}}$ (\msun)    & 55$\pm$5       & 19$\pm$2      \\
$R$ ($R_{\sun}$)&11.6$\pm$0.1   & 6.5$\pm$0.1\\
N/H          & 5 10$^{-4}$    & 6 10$^{-5}$\\
\hline
\end{tabular}
\end{center}
\end{table}

\end{document}